# Observations of Coherent Optical Transition Radiation Interference Fringes Generated by Laser Plasma Accelerator Electron Beamlets

Alex Lumpkin
*Accelerator Division*
*Fermi National Accelerator Laboratory\**, Batavia, IL USA
lumpkin@fnal,gov

Maxwell LaBerge
*Physics Department*
*Univ. of Texas-Austin\*\**
Austin, Texas USA

Donald Rule
Silver Spring, MD USA

Rafal Zgadzaj
*Physics Department*
*Univ. of Texas-Austin\*\**
Austin, TX USA

Andrea Hannasch
*Physics Department*
*Univ. of Texas-Austin\*\**
Austin, TX USA

Michael Downer
*Physics Department*
*Univ. of Texas-Austin\*\**
Austin, TX USA

Omid Zarini
*Inst. of Radiation Physics\*\*\**
*Helmholtz-Zentrum Dresden-Rossendorf,* Dresden, Germany

Brant Bowers
*Physics Department*
*Univ. of Texas-Austin\*\**
Austin, TX USA

Arie Irman
*Inst. of Radiation Physics\*\*\**
*Helmholtz-Zentrum Dresden-Rossendorf,* Dresden, Germany

Jurgen Couperus
*Inst. of Radiation Physics\*\*\**
*Helmholtz-Zentrum Dresden-Rossendorf,* Dresden, Germany

Alexander Debus
*Inst. of Radiation Physics\*\*\**
*Helmholtz-Zentrum Dresden-Rossendorf,* Dresden, Germany

Alexander Kohler
*Inst. of Radiation Physics\*\*\**
*Helmholtz-Zentrum Dresden-Rossendorf,* Dresden, Germany

Ulrich Schramm
*Inst. of Radiation Physics\*\*\**
*Helmholtz-Zentrum Dresden-Rossendorf,* Dresden, Germany

*Abstract*—We report initial observations of coherent optical transition radiation interferometry (COTRI) patterns generated by microbunched electrons from laser-driven plasma accelerators (LPAs). These are revealed in the angular distribution patterns obtained by a CCD camera with the optics focused at infinity, or the far-field, viewing a Wartski two-foil interferometer. The beam divergences deduced by comparison to results from an analytical model are sub-mrad, and they are smaller than the ensemble vertical beam divergences measured at the downstream screen of the electron spectrometer. The transverse sizes of the beamlet images were obtained with focus at the object, or near field, and were in the few-micron regime as reported by LaBerge et al. [8]. The enhancements in intensity are significant relative to incoherent optical transition radiation (OTR) enabling multiple cameras to view each shot. We present two-foil interferometry effects coherently enhanced in both the 100-TW LPA at 215 MeV energy at Helmholtz-Zentrum Dresden-Rossendorf and the PW LPA at 1.0-GeV energy at the University of Texas-Austin. A transverse emittance estimate is reported for a microbunched beamlet example generated within the plasma bubble.

## I. INTRODUCTION

Characterization of the electron beam properties in Laser-driven Plasma Accelerators (LPAs) [1] continues to be of interest as well as a challenge. This challenge includes the call for single-shot, noninvasive, and high-resolution electron beam diagnostics by the advanced accelerator community [2]. Recently, we have exploited the fact that the electron beam has been microbunched in the visible wavelength regime within the plasma bubble of the LPA as reported in the past [3-6]. This results in the generation of coherent optical transition radiation (COTR) at the surface of two thin foils located just downstream of the bubble when the beam transits them. Such effects were first observed from microbunched electrons in a self-amplified spontaneous emission (SASE) free-electron laser (FEL) experiment [7]. In that case, the increase in microbunching fraction was inherent to the exponential gain regime of the SASE FEL. The enhancements in intensity in the present case are just as significant (more than $10^4$) relative to incoherent optical transition radiation (OTR) enabling multiple cameras to view each shot. Besides the near-field imaging and use of the coherent point-spread function (PSF) for beam-size measurements reported by LaBerge et al. [8], we have also observed COTR interferometry (COTRI) fringes in the angular distribution patterns in the far-field images. These are consistent with Wartski two-foil interferometry effects [9], but coherently enhanced in both the 100-TW LPA at ~215 MeV energy at Helmholtz-Zentrum Dresden-Rossendorf (HZDR) and the PW LPA at ~1.0 GeV energy at the University of Texas (UT) at Austin. The interferences shifted the first fringe positions away from the usual single-foil $1/\gamma$ opening angle of the angular distribution pattern, and the fringe visibility indicated lower divergences for electron beamlets microbunched within narrow spectral bandwidths than the observed beam ensemble divergences. Examples from both experiments will be presented

*Work at FNAL partly supported under Contract No. DE-AC02-07CH11359 with the United States Department of Energy.
**Work by the Univ. of Texas staff supported by DoE grant DE-SC0011617 and MCD also by the Alexander von Humboldt Foundation.
*** Work at HZDR supported by the Helmholtz association under program Matter and Technology, topic Accelerator Research and Development.

as well as preliminary modeling results that elucidate the potential divergence and beam-energy information within the far-field images. A combination of the beam size and divergence data using the same narrow-band filter potentially leads uniquely to transverse emittance estimates of these microbunched beamlets generated within the plasma bubble.

## II. Experimental Aspects

### A. Facilities and Experimental Setups

The LPA experiments at HZDR used the Draco laser with a central wavelength of 800 nm, energy of up to 4 J, pulse length of 27 fs (FWHM), and peak power of 100-150 TW [10]. This laser was focused to ~20 µm (FWHM) at the center of a He gas jet (doped with 3% nitrogen) from a 3-mm diameter nozzle and resulted in plasma densities of $4.3 \times 10^{18}$ cm$^{-3}$. A laser blocking foil of 75-µm-thick Al at a few degrees off normal was located at adjustable z positions after the plasma bubble. An aluminized Kapton foil was used to block **j** x **B** electrons of low energy as well as any COTR from the back of the blocking foil and to generate the forward OTR/COTR from the aluminized back surface for the beam-size measurements. These sets of foils were loaded into a large wheel with 75 positions and changed to a new set for every shot. The OTR/COTR was redirected to a 4-cm focal length microscope objective with a working distance of 3.8 cm by a thin mirror oriented at 45 degrees to the beam direction and located 18.5 mm downstream of the first OTR foil (Fig.1). These two interfaces also formed the sources for interference fringes when viewed by the far-field (FF) focus camera. A 15-cm focal length lens was appropriately inserted in this optical path for an angular calibration factor of 0.35±0.05 mrad/pixel. The near-field (NF) focus paths had magnifications of 42 and calibration factors of 0.09 µm/pixel in the Basler 12-bit CCD cameras with 1296-pixel x 966-pixel arrays and a 3.75-µm square pixel size. Three cameras were used initially to provide single-shot images: two for orthogonal, linearly polarized COTR NF images at 600 ± 5 nm and one for the COTR FF image with no polarizer and a 633 ± 5 nm bandpass filter (BPF). Appropriate neutral density (ND) filters were added to prevent camera saturation by the enhanced radiation, and a region of interest (ROI) was identified for projected profiles.

The LPA experiments at the University of Texas (UT) at Austin used the UT-PW laser [11] with a central wavelength of 1057 nm, pulse length of ~150 fs, 100 J on target, and with a focus in the gas jet of 100 µm FWHM. The laser blocking foil and OTR source foil were changed on every shot in a vacuum-chamber access with a 1-hour turnaround time. The "x-ray imaging plate" in the spectrometer was also removed and processed after every shot. The OTR/COTR was redirected to a 10-cm focal length field lens with a working distance of <10 cm by a thin mirror oriented at 45 degrees to the beam direction and located 50.8 mm downstream of the first foil. The interferometer screen spacing was thus 50.8 mm. The near-field optical path included a prism which resulted in spectral dispersion in the horizontal plane. The FF data were obtained by removing the first field lens and placing the CCD sensor at the focal length of a 100-cm fl lens. The Peltier-cooled, 16-bit CCD camera had 5.5-µm square pixels, and the system had a spatial calibration of 0.55 µm /pixel and an angular calibration factor of 5.5 µrad/pixel.

### B. OTR and COTR Basics

Optical transition radiation (OTR) is generated by the currents induced in a material when a charged particle beam transits the interface between vacuum and the material and vice versa [12]. Both backward and forward transition radiation are generated as schematically shown in Fig. 2. The backward radiation cone of half angle 1/γ is directed around the angle of specular reflection while the forward radiation is generated in a cone around the beam direction. Thus, a 45-degree angle of the foil surface relative to the beam direction results in backward OTR at 90 degrees to the beam direction. This configuration provides a convenient access for the imaging systems to detect the radiation and image the beam distribution. In consideration of Fig. 2a and Fig. 2b, one notes that forward OTR from screen 1 could arrive at the second foil and interfere with the backward OTR from screen 2 to form fringes in the angular distribution pattern. This occurs when the foil separation, L, is of order γ²λ, where γ is the relativistic Lorentz factor for the electron beam and λ is the observation wavelength. We take advantage of this aspect in our experiments.

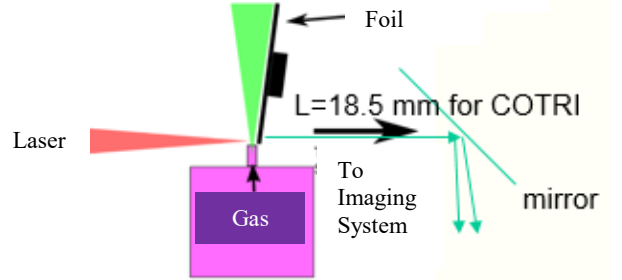

Fig. 1. Schematic of the HZDR setup with laser, gas jet, foil wheel, mirror, and COTRI source setup. The microscope objective is focused on the thin foil after the blocking foil. The foil spacing L=18.5 mm is shown.

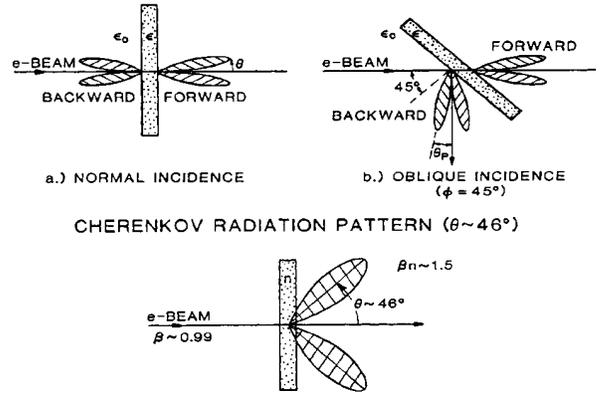

Fig. 2. Schematic of OTR sources at the interfaces of vacuum and a material with dielectric constant ε for a) normal incidence, b) oblique incidence, 45-degrees, and c) a comparison to the Cherenkov radiation pattern with an opening angle of ~46 degrees from fused silica [13].

The single electron OTR spectral angular distribution of the number of photons per unit frequency per steradian is given by (1), where ω is the angular frequency, Ω is the solid angle, Planck's constant/2π is ℏ, e is the electron charge, c is the speed of light, and $\theta_x$ and $\theta_y$ are radiation angles [14]. This results in the well-known OTR angular distribution pattern with the $1/\gamma$ opening angle as illustrated in Fig. 3. This case is for 220 MeV with divergences in both planes of 0.2 mrad, and the opening angle is 2.3 mrad.

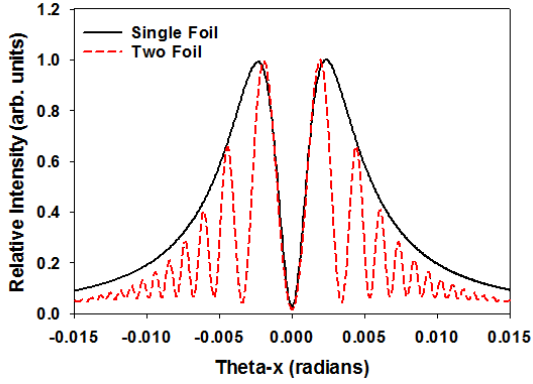

Fig. 3. Analytical model calculations for the OTR angular distribution patterns as seen in FF imaging for 220 MeV and divergences of $\sigma_{x',y'}$ = 0.2 mrad: single-foil source and Wartski two-foil interferometer with L= 6.3 cm and λ=537 nm. The single foil lobes at 2.3 mrad show the beam energy effect, and the central minimum is divergence dependent while the two-foil outer interference fringes are more divergence sensitive at low values than the single foil curve.

$$\frac{d^2W_1}{d\omega d\Omega} = \frac{e^2}{\hbar c} \frac{1}{\pi^2 \omega} \frac{(\theta_x^2 + \theta_y^2)}{(\gamma^{-2} + \theta_x^2 + \theta_y^2)^2} \quad (1)$$

The incorporation of OTR interference and coherence terms are addressed in (2)-(4) and produce the full spectral angular distribution function for $N$ particles. Equation (2) shows the separated functions that modify the single-electron function shown above:

$$\frac{d^2W}{d\omega d\Omega} = |r_{\|,\perp}|^2 \frac{d^2W_1}{d\omega d\Omega} I(k) J(k) \quad (2)$$

where the reflection coefficient components are $r_{\|,\perp}$, the interference function is $I(k)$, and the coherence function is $J(k)$. The interference function is given in (3) where $L$ is the foil separation distance, $k = |k| = 2\pi/\lambda$ and with the assumed small-angle approximations,

$$I(k) = 4 \sin^2\left[\frac{kL}{4}\right] (\gamma^{-2} + \theta_x^2 + \theta_y^2) \quad (3)$$

The corresponding two-foil plot is also shown in Fig. 3 for L= 6.3 cm and λ= 537 nm. Strong fringe modulation is seen in this example where a Gaussian beam divergence of 0.2 mrad was convolved with (2).

The coherence function can be defined in (4) with a microbunching fraction $f_B = N_B/N$ where $N_B$ is the number of microbunched electrons and $H(k)$ is the Fourier transform of the charge form factors. Note the coherence function reduces just

$$J(k) = N + N_B(N_B - 1)|H(k)|^2 \quad (4)$$
$$\text{where} \quad H(k) = \frac{\rho(k)}{Q} = g_x(k_x) g_y(k_y) F_z(k_z).$$

to the number of particles, $N$, when the number of microbunched particles, $N_B$, is zero. Q is the total charge, and $g(k_i)=\exp(-\sigma_i^2 k_i^2/2)$ for $i=x,y$ are the transverse charge form factors, with $k_i \approx k\theta_i$. Here, $F_z(k_z)=\exp(-\sigma_z^2 k_z^2/2)$ is the Fourier transform of the longitudinal form factor with $k_z \sim k$ for $\theta \ll 1$.

### III. EXPERIMENTAL AND ANALYTICAL RESULTS

Interesting imaging results have been obtained at both facilities at about 215 MeV and 1.0 GeV, respectively. The data are compared to COTRI analytical model results. Previously we had suggested that incoherent OTR techniques would be applicable [15,16], but the coherent enhancements dominate at these wavelengths for these LPA microbunched electrons.

#### A. HZDR LPA Results

The first experimental results evaluated were at a beam energy of 215 MeV with 15-MeV FWHM energy spread and ~100 pC in the quasi-monoenergetic peak. The FF image is shown in Fig. 4a with opening angles at ±4.75 mrad and with extensive fringes (6 in the image view, more at larger angles) in positive $\theta_y$ and only the first peak strongly enhanced in negative $\theta_y$ angles. The $\theta_x$ fringes are also limited suggesting a larger beam size whose Fourier transform does not enhance the fringes at larger angles. The fringe visibility is assessed by comparison to the analytical results in Fig. 5 for 0.5-mrad and 1.0-mrad divergence cases, and the e-beam divergence is consistent with the 0.5-mrad case, or lower, based on fringe visibility. This value is smaller than the ensemble divergence measured at the spectrometer, although foil scattering terms need to be considered. When this 0.5-mrad y divergence estimate is combined with the beam-size measurement of $\sigma_y$=1.5 μm on the same shot (but slightly different wavelength), an estimated normalized emittance of $\varepsilon_n$~0.32 mm-mrad (rms) for the microbunched beamlet is obtained. This is the first demonstration of the COTR single-shot emittance concept on an LPA.

Additionally, we estimated a COTR/OTR gain of about $10^5$ since the data involved a 30 times narrower 10-nm wide sampling (instead of integrating the visible light region from 400-700 nm), an ND 2.6 filter with attenuation of 400 was used, and a 10 times lower charge of only 100 pC was used compared to a few nC normally needed for OTRI imaging with a CCD camera (depending on the angular magnification). This gain implies significant microbunching at the 1 to 2% level occurred.

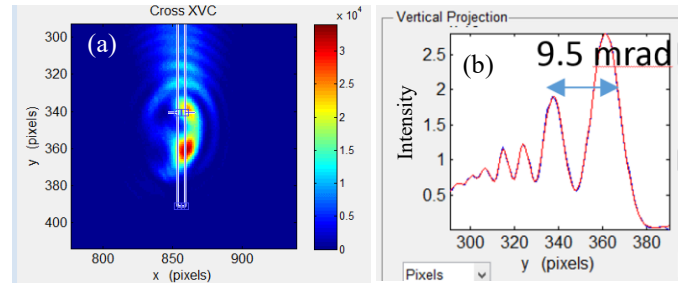

Fig. 4. (a) Initial COTR angular distribution pattern and (b) vertical projection of the ROI for a 215-MeV quasi-monoenergetic beam showing a strong fringe pattern in positive $\theta_y$ angles at 633 nm. The data (blue) were matched with a curve composed of seven single-Gaussian peaks (red) to locate fringe positions.

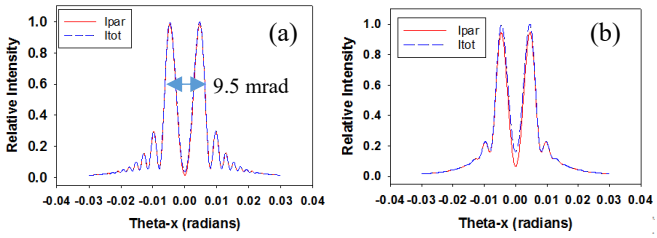

Fig. 5. COTRI analytical results for divergences of a) 0.5 mrad and b) 1.0 mrad with E=215 MeV, λ=633±5 nm, and $\sigma_{x,y}$= 2 µm beam sizes. The parallel polarization and total components, $I_{par}$ and $I_{tot}$, respectively should be compared.

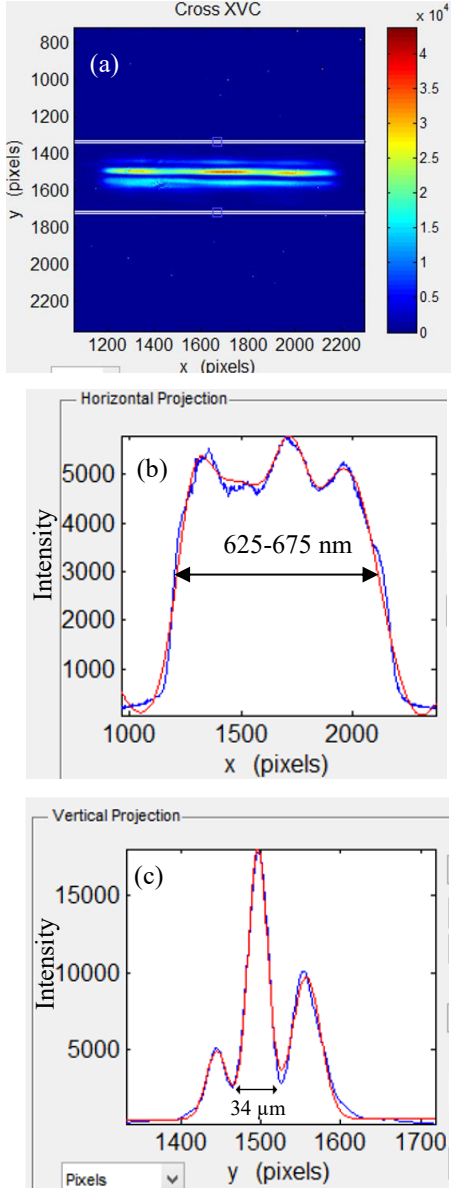

Fig. 6. (a) Near-Field image with vertical polarizer used, (b) horizontal projected profile. and (c) vertical projected profile for the ROI from the UT-PW LPA 2017 run, Shot #10830R. The prism provided spectral dispersion on the horizontal axis within the 625-675 nm BPF. The vertical projection in c) is consistent with two beamlets separated vertically by ~34 µm whose coherent OTR PSF lobes overlap in the middle stronger peak. The data (blue) are matched by a 5-Gaussian peak fit (red) and a 3-Gaussian peak fit (red) in (b) and (c), respectively.

## B. UT-PW LPA Results

Only a limited number of shots on two runs were obtained on the PW LPA, but there were several interesting features observed including the wavelength dispersion from the prism in one axis with a BPF employed from 625-675 nm. An example of such a NF image is shown in Fig. 6a. In this case, the horizontal display axis is the wavelength-dispersive axis and the vertical display axis is the y-spatial axis as shown in the projections below the image. One notes the intensity modulation in the image in the horizontal projection which is attributed to COTR within the BPF span. Incoherent OTR does not exhibit such structures. The vertical projection is consistent with two beamlets whose coherent PSF model lobes overlap in the middle stronger peak. Using three Gaussian single peaks to fit the profiles, the PSF lobe separations are 51 pixels and 60.5 pixels left and right, respectively, which imply beam sizes of about 28 µm and 34 µm FWHM (using the scaling from the coherent PSF model lobe separation plot at 600 nm and adjusting for 650 nm [8]). The beamlet separations are 34 µm based on the two observed presumed minima, and the right beamlet is noticeably 2-3 times stronger in intensity than the left one.

Another aspect from the same run involved a FF imaging setup with the first lens removed, a lens located at a distance equal to its focal length, and the same high-sensitivity camera. An example image is shown in Fig. 7 with a vertical polarizer and the prism still installed for dispersion. We had anticipated a single-foil-like result with opening lobes of 1/γ = ± 0.50 mrad, but the interference effects dominated and shifted the first peaks outward to ~ ± 2.8 mrad. Our COTRI model indicated this would be consistent with L= 50.8 mm, E = 1.0 GeV, and λ=650 ± 25 nm as shown in Fig. 8a. The beam energy was independently measured in the downstream electron spectrometer with a semi-quasi-monoenergetic peak near 1 GeV. The acceptance angle for the FF imaging covered about 15 mrad total in the $\theta_y$ axis, and the second fringe peaks are calculated at ~ ± 6 mrad. The absence of the second peak in -$\theta_y$ angles is attributed to the relatively small coherence factor at larger angles for a 30-µm microbunched beam size compared to a 10-µm beam size as illustrated in Fig. 8a. A 50-µrad divergence value was used in the model. In Fig. 8b we show the effect of beam divergence on the outer peak visibility for 100 and 700 µrad. The latter has reduced fringe modulation and could be another factor reducing peak 2.

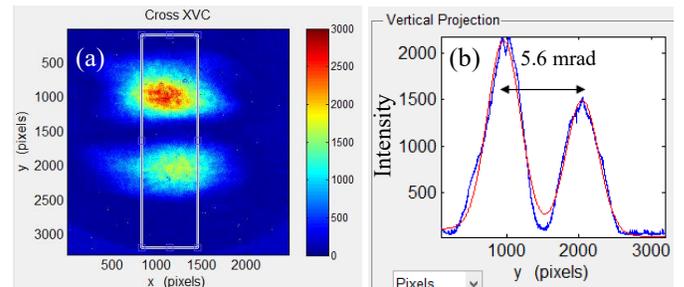

Fig. 7. (a) Far field image with vertical polarizer used and (b) vertical projection in ROI for UT-PW LPA run, Shot # 10858R. Single Gaussian peak fits (red) and data (blue) are shown. The first fringes are not at ± 0.50 mrad which is 1/γ for 1.0-GeV beam, but they are at ~ ± 2.8 mrad which is consistent with interference effects for L=50.8 mm and λ=650 ± 25 nm.

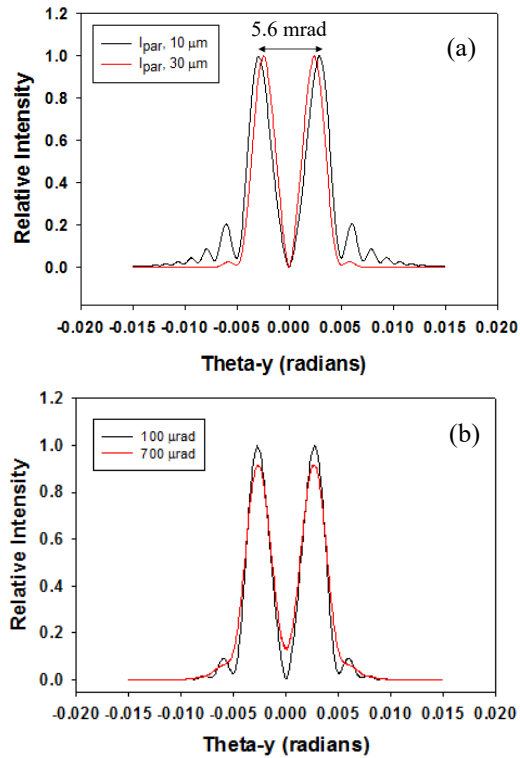

Fig. 8. Plots of COTRI patterns at 1.0 GeV for (a) two beam sizes showing the coherence gain factors on the observed fringes for the parameters in Fig. 7. Peaks at angles beyond ±5 mrad have lower gain factors in the $\sigma_y$ = 30-μm case. and (b) two divergence values showing the effect on the central minimum and peaks 2,3 with reduced visibility at the 700-μrad case with $\sigma_y$ = 20 μm.

Since LPA electron bunches are strongly divergent as well as ultrashort in duration, it may be necessary to account for divergence immediately when summing/integrating the COTR fields, as is done in [17], rather than after determining intensities, as is presented here. A model following the former approach may make different estimates for the divergence of these coherent electron bunches. Further analyses will be done to determine if such an approach is necessary for LPA beams. The model presented here represents a preliminary attempt to determine LPA beam emittance estimates directly from COTR.

## IV. Summary

In summary, we have identified significant microbunching in the electron beamlets accelerated in LPAs at 215 MeV and 1.0 GeV with concomitant COTR signal enhancements. Using far-field imaging, we have also identified COTR interference fringes and matched the data features with the COTRI model [14]. This analytical model, initially developed for the SASE-FEL-induced microbunching in the visible regime, has been applied to the LPA cases for the first time. We note the implied divergences of the microbunched beamlets are less than those of the ensemble of electrons measured at the spectrometer screen, in part because beam scattering by the turning mirror does not affect the COTRI. A single-shot emittance estimate was reported for a microbunched beamlet. These effects warrant further investigations and simulations to elucidate the fundamental LPA process in the bubble regime.

## Acknowledgments

The authors acknowledge the Draco laser staff at HZDR and the PW laser staff at UT-Austin for operating the high-power laser systems for these LPA experiments.